\documentclass[12pt,a4paper]{article}
\usepackage{jheppub}
\pdfoutput=1
\usepackage{caption}
\usepackage{dsfont}
\usepackage{amssymb,amsmath}
\usepackage{epsfig}
\usepackage{epstopdf}
\usepackage{latexsym}
\usepackage{graphicx}
\usepackage{subfigure}
\usepackage{booktabs}
\usepackage{bbm}
\makeatletter
\def\@fpheader{\relax}
\makeatother

\newcommand{\be}{\begin{equation}}
\newcommand{\ee}{\end{equation}}
\newcommand{\bea}{\begin{eqnarray}}
\newcommand{\eea}{\end{eqnarray}}

\subheader{\begin{flushright}
UTTG-18-15 \\
TCC-007-15
\end{flushright}}

\title{An equal area law for holographic entanglement entropy of the AdS-RN black hole}
\author{Phuc H. Nguyen}
\affiliation{Department of Physics, University of Texas at Austin,\\2515 Speedway, Austin, TX 78712-1192, U.S.A.}
\emailAdd{phn229@physics.utexas.edu}

\abstract{The Anti-de Sitter-Reissner-Nordstr\"{o}m (AdS-RN) black hole in the canonical ensemble undergoes a phase transition similar to the liquid-gas phase transition, i.e. the isocharges on the entropy-temperature plane develop an unstable branch when the charge is smaller than a critical value. It was later discovered that the isocharges on the {\it entanglement entropy}-temperature plane also exhibit the same van der Waals-like structure, for spherical entangling regions. In this paper, we present numerical results which sharpen this similarity between entanglement entropy and black hole entropy, by showing that both of these entropies obey Maxwell's equal area law to an accuracy of around 1 \%. Moreover, we checked this for a wide range of size of the spherical entangling region, and the equal area law holds independently of the size. We also checked the equal area law for AdS-RN in 4 and 5 dimensions, so the conclusion is not specific to a particular dimension. Finally, we repeated the same procedure for a similar, van der Waals-like transition of the dyonic black hole in AdS in a mixed ensemble (fixed electric potential and fixed magnetic charge), and showed that the equal area law is not valid in this case. Thus the equal area law for entanglement entropy seems to be specific to the AdS-RN background.}
\begin{document}

\maketitle
\flushbottom

\section{Introduction}\label{Sec:intro}
Entanglement entropy (EE) appears to be a versatile tool that can be used to study a rich variety of physical phenomena. In particular, it can serve as a probe of the different phases of the theory \cite{Calabrese:2004eu,Calabrese:2005zw}, ranging from the confining phase of large-$N$ gauge theories \cite{Klebanov:2007ws} to topological phases in condensed matter systems \cite{Kitaev:2005dm}, to tachyon condensation \cite{Nishioka:2006gr} and superconducting phase transitions \cite{Albash:2012pd}.

Entanglement entropy has also emerged as a central component of the AdS/CFT correspondence. According to the Ryu-Takayanagi formula \cite{Ryu:2006bv,Nishioka:2009un}, the entanglement entropy $S_{A}$ between a boundary region $A$ and its complement is computed (in static backgrounds) in an elegant geometric fashion as the area of a minimal surface. The striking similarity between the Ryu-Takayanagi formula and the Bekenstein-Hawking formula for black hole entropy suggests some deep connection between entanglement entropy and black hole entropy. It has even been suggested that the origin of black hole entropy is entanglement entropy \cite{Srednicki:1993im, Frolov:1993ym, Solodukhin:2011gn}.

Motivated by the themes above, in this paper we track entanglement entropy across a family of van der Waals-like phase transitions of charged black holes in AdS. The first phase transition under study is the one of AdS-RN in the canonical ensemble in 4 and 5 dimensions. This transition was first discovered in \cite{Chamblin:1999tk,Chamblin:1999hg}: the curves of constant charge on the temperature-entropy plane have an unstable portion when the charge is smaller than a critical value. Moreover, at critical charge, the unstable portion squeezes to an inflection point. It was subsequently pointed out in \cite{Johnson:2013dka} that the same qualitative behavior can be observed if we study the isocharges in the {\it entanglement entropy}-temperature plane (for spherical entangling regions), and that moreover the inflection point occurs at the same critical temperature as the black hole with the same critical exponent. These findings were then generalized to a wider class of supergravity backgrounds in \cite{Caceres:2015vsa}, where it was found that, in all cases, the isocharges on the {\it entanglement entropy}-temperature plane mimick the qualitative behavior of the ones on the entropy-temperature plane.

In this paper, we ask the question of how far we can push this similarity between the two kinds of entropies with respect to these phase transitions. In particular, we will investigate whether Maxwell's equal area construction, which is known to hold in temperature-entropy plane, is also valid in the temperature-entanglement entropy plane. Maxwell's construction in the context of black hole thermodynamics has generated some recent interest: this topic was studied in \cite{Spallucci:2013osa, Spallucci:2013jja} in the context of AdS-RN, but also in \cite{Belhaj:2014eha,Lan:2015bia} in other contexts. We will show that, for AdS-RN, the van der Waals behavior on the entanglement entropy-temperature plane indeed also obeys the equal area law, with the transition temperature obtained by minimizing the black hole free energy function. We also vary the size of the spherical entangling region over a wide range and checked that the conclusion is valid to an excellent accuracy. Moreover, we checked this for the 4-dimensional and 5-dimensional black holes, so that the equal area law is not specific to a particular dimension. We also repeated the same procedure for another van der Waals-like phase transition: that of the dyonic black hole in the fixed electric potential and fixed magnetic charge ensemble, and our numerical results indicate that the equal area law does not hold in this case. Thus, the equal area law seems to be a specificity of the AdS-RN background.

The rest of the paper is organized as follows: In Section \ref{Sec:AdS4}, we review the phase structure of AdS-RN in 3+1 dimensions in the canonical ensemble and discuss Maxwell's equal area law in the entropy-temperature plane. In Section \ref{Sec:MaxwellAdSRN}, we then turn to the numerical computation of holographic entanglement entropy, and present numerical evidence that the equal area construction also holds on the entanglement entropy-temperature plane. In Section \ref{Sec:DyonicAdS4}, we repeat for the dyonic black hole in 3+1 dimensions and show that the equal area construction is not valid for this background. Next, in Section \ref{AdS5}, we check that the equal area law also holds for AdS-RN in 4+1 dimensions. Finally, in Section \ref{Sec:Conclusion}, we summarize our main findings and discuss a few possible future work.

\section{Review of 4d AdS-RN in the canonical ensemble}\label{Sec:AdS4}
\begin{figure}
$$
\begin{array}{cc}
 \includegraphics[width=7.5cm]{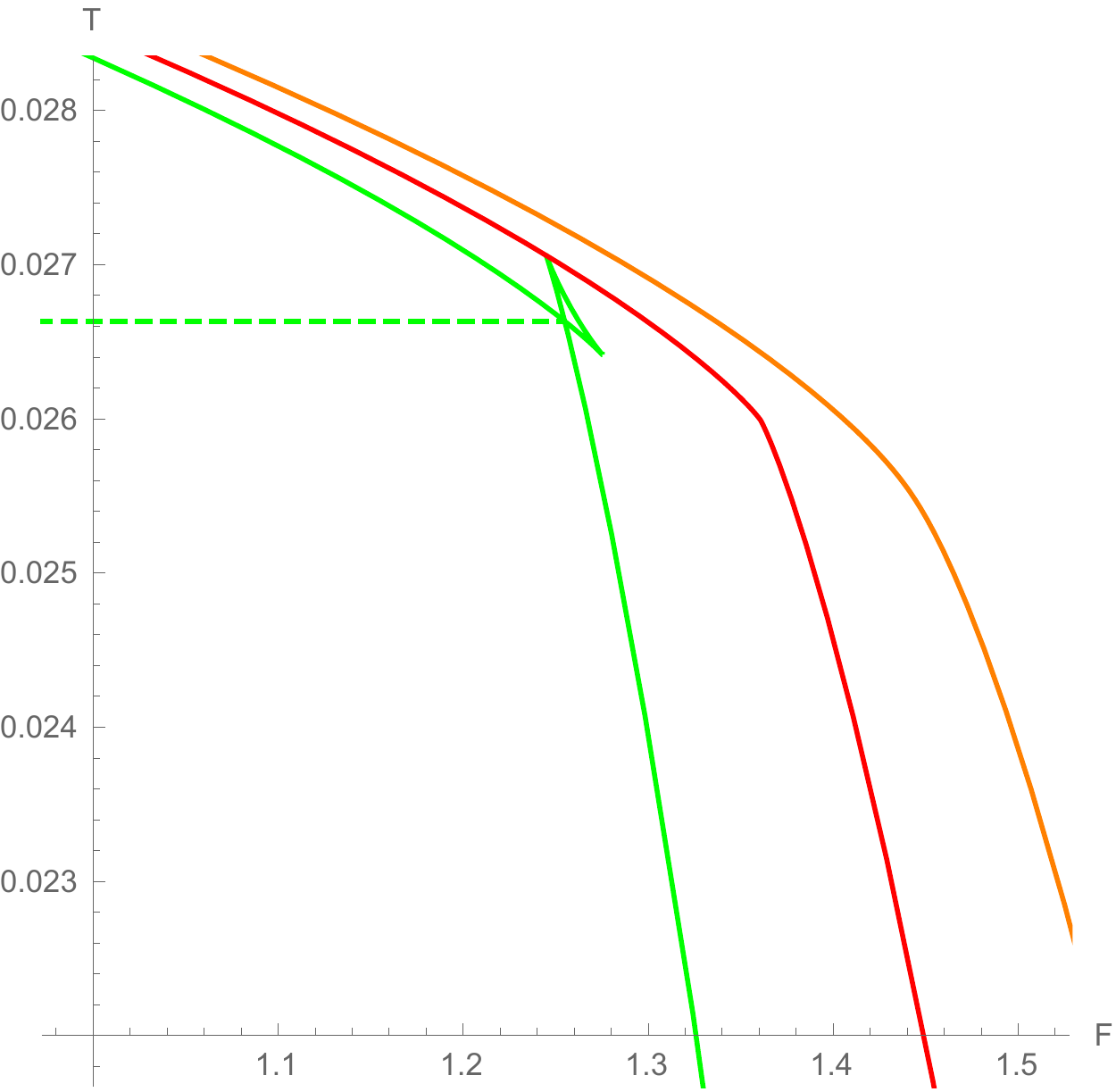} & \includegraphics[width=7.5cm]{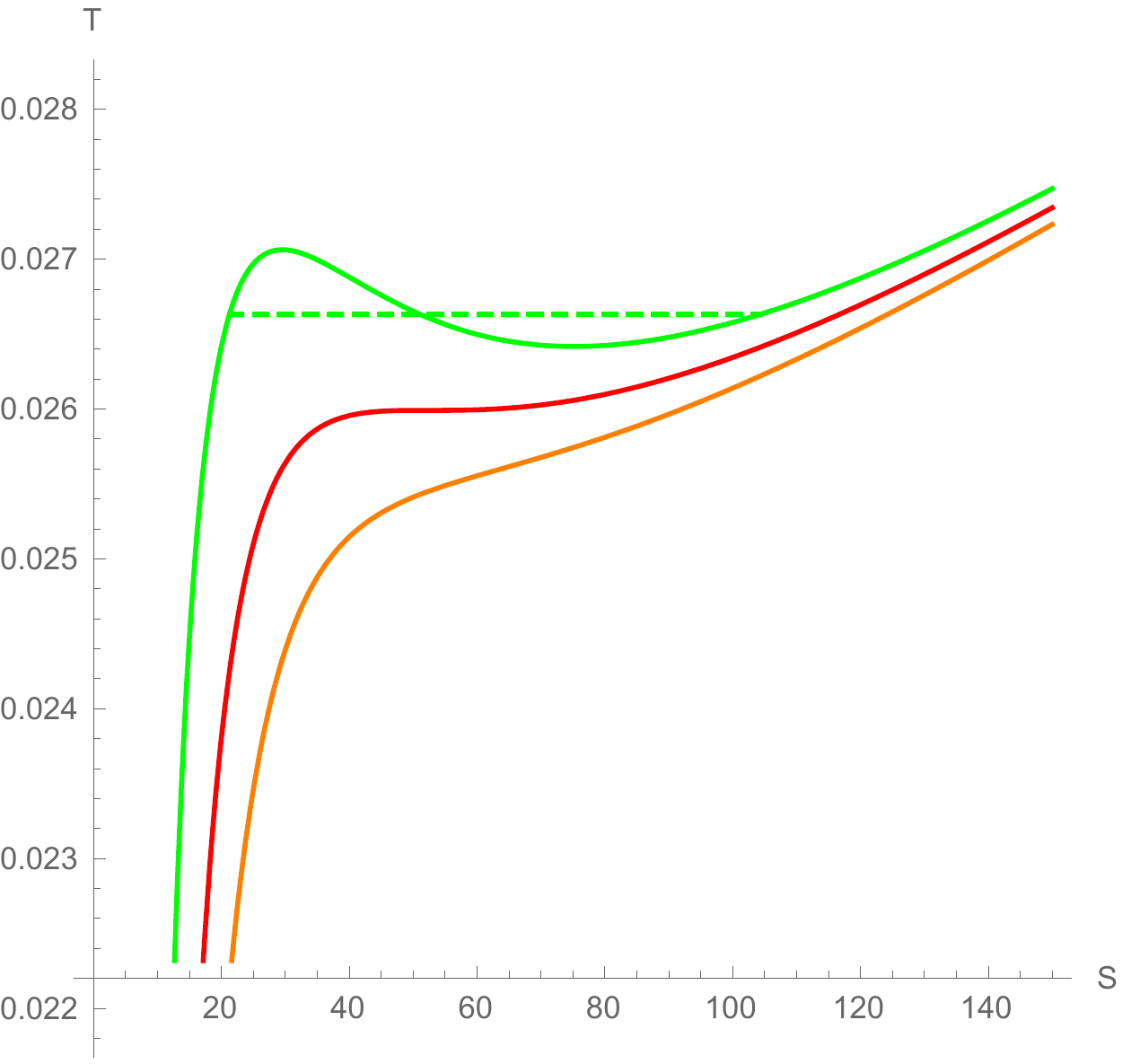}\\
\end{array}
$$
\caption{Plot of the temperature versus the free energy (left panel) and plot of the temperature versus the entropy (right panel). For both panels, the values of the charge are $Q=1.5$ (green), $Q=Q_{c}=\frac{10}{6}$ (red) and $Q=1.8$ (orange), and we use $L=10$. The transition temperature for the green curve is $T_{*} = 0.02663$.}
\label{AdSRNPlot}
\end{figure}

In this section, we survey the phase structure of the 4-dimensional AdS-RN black hole in the fixed charge ensemble, leading up to the van der Waals behavior in the entropy-temperature plane (i.e. there exists a family of first order transition ending with a second order one) and Maxwell's construction \footnote{The distinction between ``first order'' and ``second order'' here refers to the slope of the free energy plot versus the temperature. Upon a closer look, the nature of these phase transitions is more subtle: see for example \cite{Banerjee:2010da} where the phase transition was studied using Ehrenfest equations.}. The Einstein-Maxwell action in 4 dimensions reads:
\begin{equation}\label{EinsteinMaxwell}
I = -\frac{1}{16\pi} \int d^{4}x \sqrt{-g} (R-2\Lambda-F^{2})\,.
\end{equation}
The AdS-RN solution together with the gauge field is given by:
\begin{equation}
ds^{2} = -f{(r)}dt^{2} + \frac{dr^{2}}{f{(r)}} + r^{2}(d\theta^{2}+\sin^{2}{\theta}d\phi^{2})\,,
\end{equation}
\begin{equation}
f{(r)} = 1 - \frac{2M}{r} + \frac{Q^{2}}{r^{2}} + \frac{r^{2}}{L^{2}}\,,
\end{equation}
\begin{equation}
A = Q\left(\frac{1}{r_{+}}-\frac{1}{r}\right)dt\,.
\end{equation}
where $M$ is the mass, $Q$ is the electric charge and $L$ is the AdS lengthscale. The additive constant in $A_{t}$ was chosen to be $\frac{Q}{r_{+}}$ so that the norm of the vector potential $A^{2}$ is regular at the horizon. The black hole temperature and entropy are:
\begin{equation}\label{T4d}
T = \frac{3r_{+}^{4}+L^{2}(r_{+}^{2}-Q^{2})}{4\pi L^{2}r_{+}^{3}}\,,
\end{equation}
\begin{equation}\label{S4d}
S = \pi r_{+}^{2}\,,
\end{equation}
where $r_{+}$ is the horizon (the largest root of $f{(r_{+})}=0$). From (\ref{T4d}) and (\ref{S4d}), we can easily eliminate the parameter $r_{+}$ to obtain the function $T(S,Q)$:
\begin{equation}
T{(S,Q)} = \frac{1}{4\pi}\left(\frac{3}{L^{2}}\sqrt{\frac{S}{\pi}} + \sqrt{\frac{\pi}{S}} - Q^{2}\frac{\pi^{3/2}}{S^{3/2}} \right)\,.
\end{equation}
From the function $T(S,Q)$ above, one can plot the isocharges on the $T-S$ plane. The plot is presented on the right panel of Figure \ref{AdSRNPlot}. As can be seen from the plot, the curve is monotonic for sufficiently large $Q$. As $Q$ decreases, the curve has an inflection point when $Q$ reaches a threshold value $Q_{c}$. One can solve for the position of the inflection point by:
\begin{equation}
\left(\frac{\partial T}{\partial S}\right)_{Q} = \left(\frac{\partial^{2} T}{\partial S^{2}}\right)_{Q} = 0\,.
\end{equation}
We find the critical entropy $S_{c}$, critical charge $Q_{c}$ and critical temperature $T_{c}$ to be:
\begin{equation}
S_{c} = \frac{\pi}{6}L^{2}\,,
\end{equation}
\begin{equation}
Q_{c} = \frac{L}{6}\,,
\end{equation}
\begin{equation}
T_{c} = \sqrt{\frac{2}{3}}\frac{1}{\pi L}\,.
\end{equation}
Finally, when $Q < Q_{c}$, the curve becomes oscillatory, and there is a small portion with negative heat capacity:
\begin{equation}
T\left(\frac{\partial S}{\partial T}\right)_{Q} \leq 0\,.
\end{equation}
Like in the case of the liquid-gas transition, this portion is believed to be thermodynamically unstable, and should be replaced by an isotherm $T = T_{*}$ according to Maxwell's prescription. The value of $T_{*}$ can be obtained in two different (but equivalent) ways: by the equal area condition, or from the Helmholtz free energy. The first method, the equal area condition, states that $T_{*}$ is the unique temperature which divides the oscillatory part of the curve $T(S)$ into two regions with equal area. We will find $T_{*}$ from the second method, i.e. using the Helmholtz free energy, and check numerically that it is equivalent to the first. The Helmholtz free energy can be found from the on-shell action \footnote{The Helmholtz free energy of AdS-RN is usually measured with respect to an extremal background with the same electric charge \cite{Chamblin:1999tk}. For our purposes, however, this background subtraction only shifts the plot of $F$ versus $T$ in the vertical direction and does not affect the transition temperature $T_{*}$.}:
\begin{equation}\label{Helmhotz}
F = \frac{1}{4L^{2}}\left(L^{2}r_{+}-r_{+}^{3}+\frac{3Q^{2}L^{2}}{r_{+}}\right)\,.
\end{equation}
We present in the left panel of Figure \ref{AdSRNPlot} the plot of $F$ versus $T$. For $Q < Q_{c}$, we observe the swallowtail behavior familiar from catastrophe theory, and the transition temperature $T_{*}$ is the horizontal coordinate of the junction between the two stable branches. Numerically, using $Q=1.5$ and $L=10$, we found $T_{*} \approx 0.02663$.
We can now proceed to verify Maxwell's equal area law, which is the following statement:
\begin{equation}\label{MaxwellLaw}
\int_{S_{1}}^{S_{2}} T{(S,Q)} dS - T_{*}(S_{2}-S_{1}) = T_{*}(S_{3}-S_{2}) - \int_{S_{2}}^{S_{3}} T{(S,Q)} dS
\end{equation}
where $S_{1}$, $S_{2}$ and $S_{3}$ are the smallest, intermediate, and largest roots of the equation $T_{*} = T{(S,Q)}$. Numerically, we found that both sides of (\ref{MaxwellLaw}) evaluate to around 0.00761, thus confirming the validity of Maxwell's construction.

\section{Maxwell construction for entanglement entropy}\label{Sec:MaxwellAdSRN}
In this section, we investigate the phase structure of the EE-temperature plane instead of the entropy-temperature plane. As discussed in the introduction, the work of \cite{Johnson:2013dka} demonstrates that, at least for AdS-RN in 4d, the isocharges on the EE-temperature plane mimick the qualitative behavior of the entropy-temperature plane. However, that paper did not make the connection with Maxwell's construction. In this section, we will show that, indeed, entanglement entropy also respects the Maxwell construction, thereby strengthening the conclusion of \cite{Johnson:2013dka}. First, let us briefly review a few field theory generalities about EE.

\subsection{Brief review of holographic EE}\label{EEsubsec}
Suppose we have a quantum field theory described by a density matrix $\rho$, and let $A$ be some region of a Cauchy surface of spacetime. The entanglement entropy between $A$ and its complement $A^{c}$ is defined to be:
\begin{equation}
S_{A} = -\mathrm{Tr}_{A}{(\rho_{A}\log{\rho_{A}})}\,,
\end{equation}
where $\rho_{A}$ is the reduced density matrix of $A$: $\rho_{A}=\mathrm{Tr}_{A^{c}}{(\rho)}$. As mentioned in the introduction, entanglement entropy is computed holographically by the Ryu-Takayanagi recipe \footnote{The Ryu-Takayanagi formula only applies to static backgrounds and when the bulk theory is Einstein gravity.}:
\begin{equation}
S_{A} = \frac{\text{Area}(\Gamma_A)}{4 G_N}\,,
\end{equation}
where $\Gamma_A$ is a codimension-2 minimal surface with boundary condition $\partial \Gamma_A=\partial A$, and $G_{N}$ is the gravitational Newton's constant.

A few remarks are in order concerning EE at finite temperature and in finite volume. At nonzero temperature, entanglement entropy no longer has the nice properties that it does at zero temperature, for example the area law (see, for example, \cite{Fischler:2012ca}). This is due to the fact that, at finite temperature, entanglement entropy is ``contaminated'' by a thermal component which scales as the volume of the entangling region rather than its area.

Moreover, when the bulk is topologically nontrivial (as is the case for AdS-RN with spherical horizon), the Ryu-Takayanagi formula is refined by an additional topological constraint: only surfaces which are homological to the entangling region on the boundary are considered in the minimization problem \cite{Hubeny:2013gta}. This constraint, which ensures that the Araki-Lieb inequality is satisfied, implies that, for sufficiently large entangling region, the minimal surface is disconnected and includes the horizon itself as a connected component. It might appear curious that the homology constraint means the entanglement entropy of $A$ is not equal to that of the complement of $A$, but this is expected when the system is in a mixed state. By avoiding large entangling regions, we will also avoid having to deal with this ``phase transition'' between connected and disconnected minimal surfaces \footnote{Similar phase transitions are also observed in infinite volume, when the entangling region is composed of multiple strips \cite{Ben-Ami:2014gsa}.}.

\subsection{Maxwell's equal area law}\label{Subsec:MaxwellEE}
\begin{figure}
$$
\begin{array}{cc}
 \includegraphics[width=7.5cm]{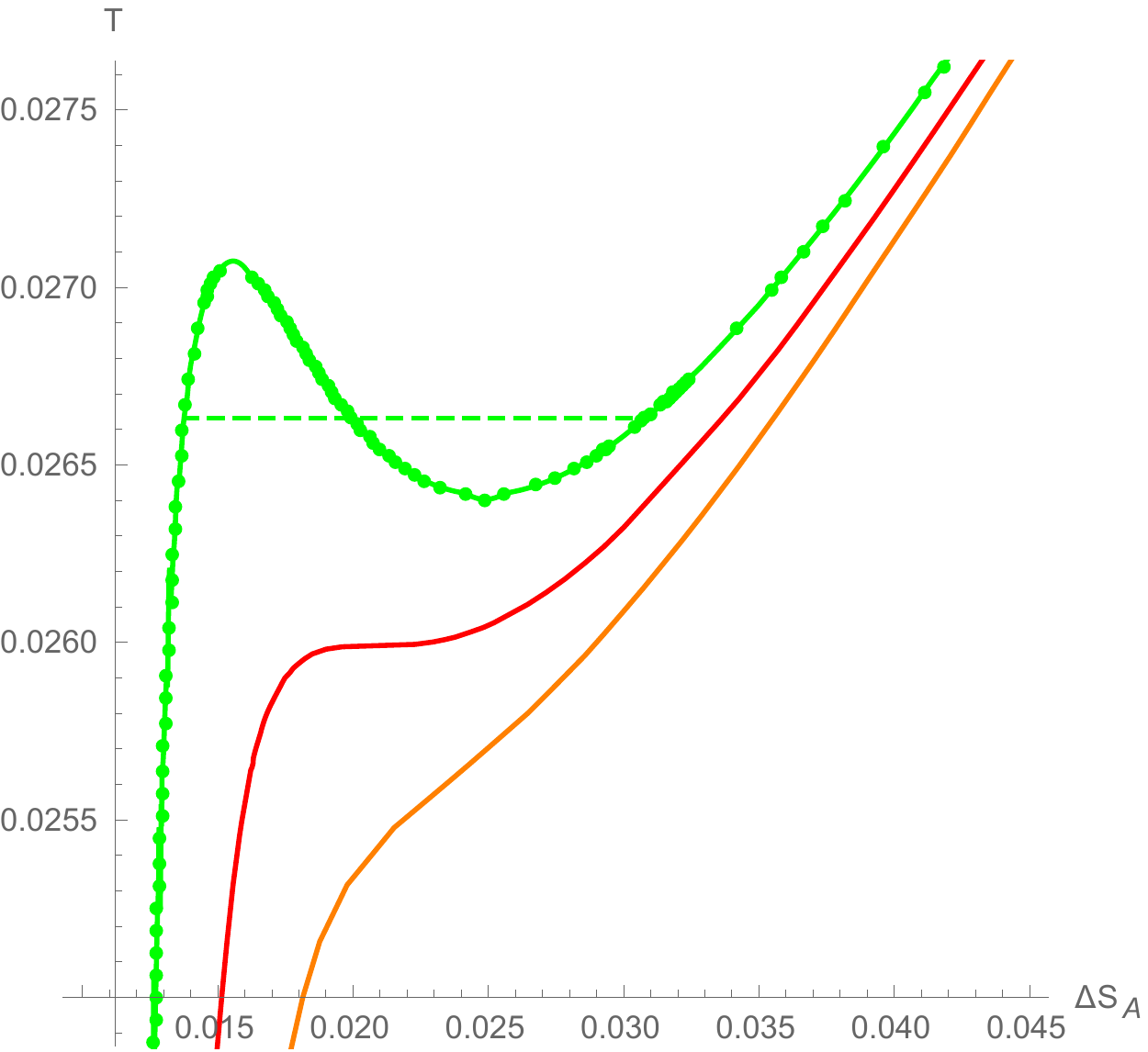} & \includegraphics[width=7.5cm]{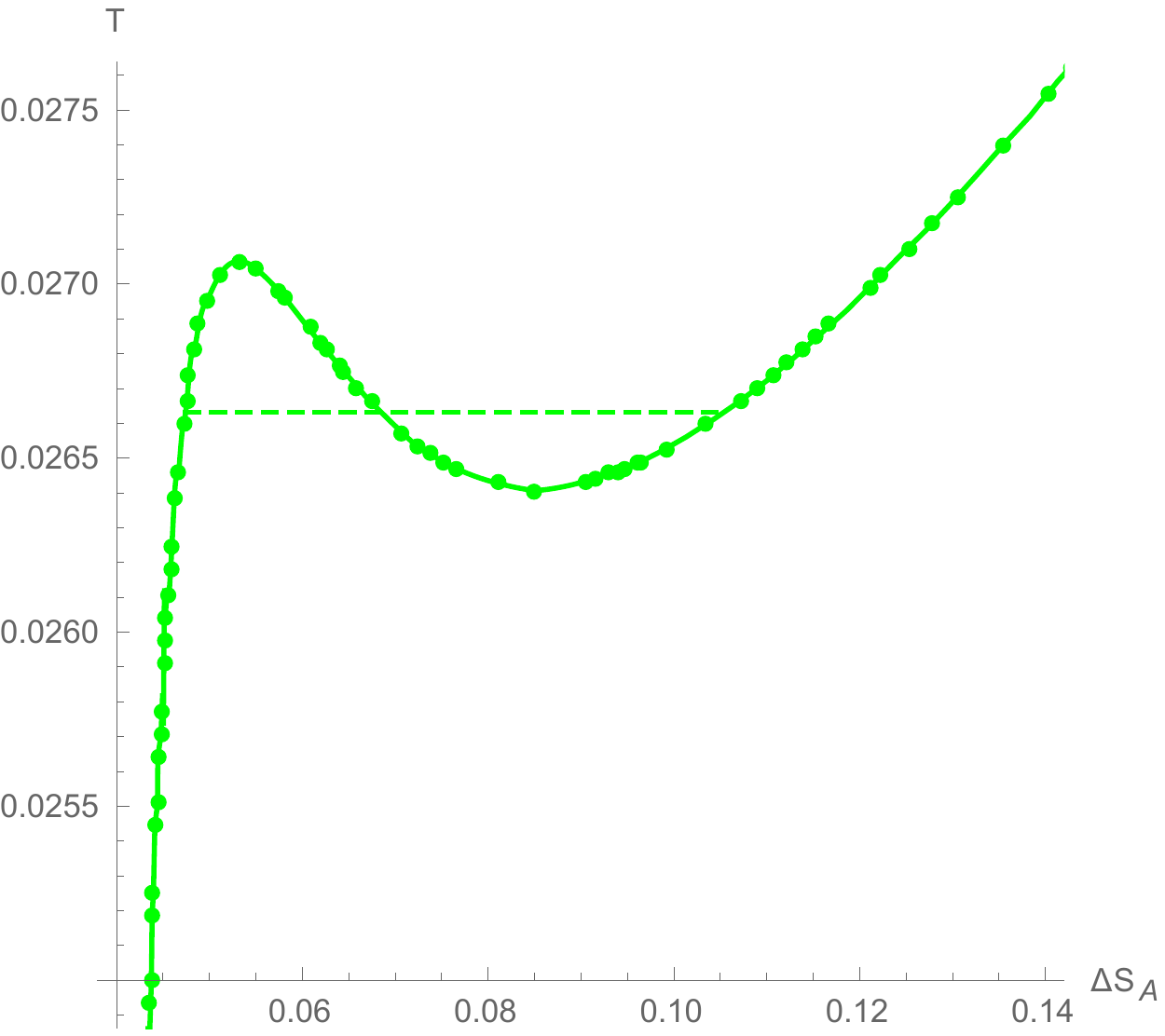}\\
  \includegraphics[width=7.5cm]{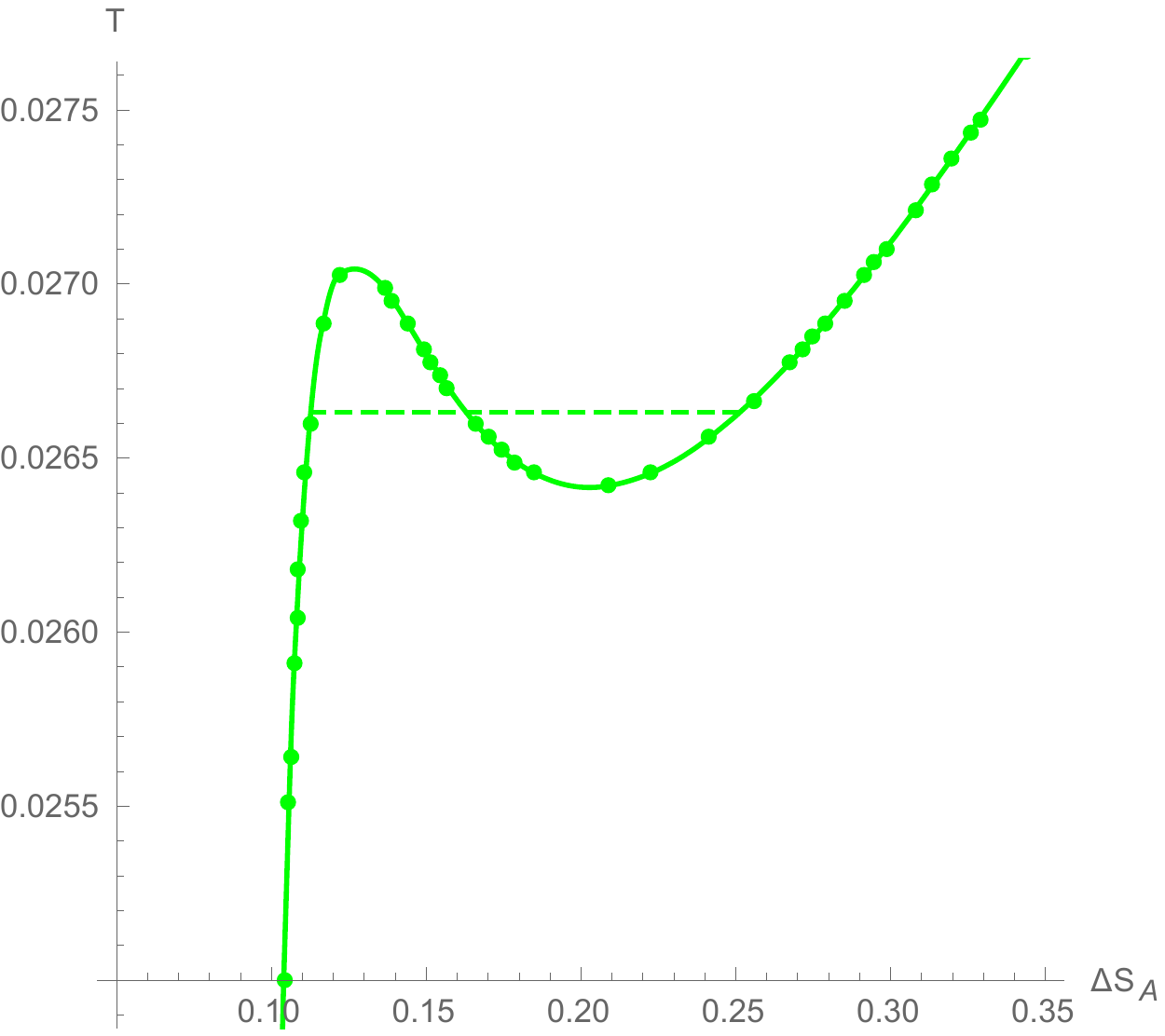} & \includegraphics[width=7.5cm]{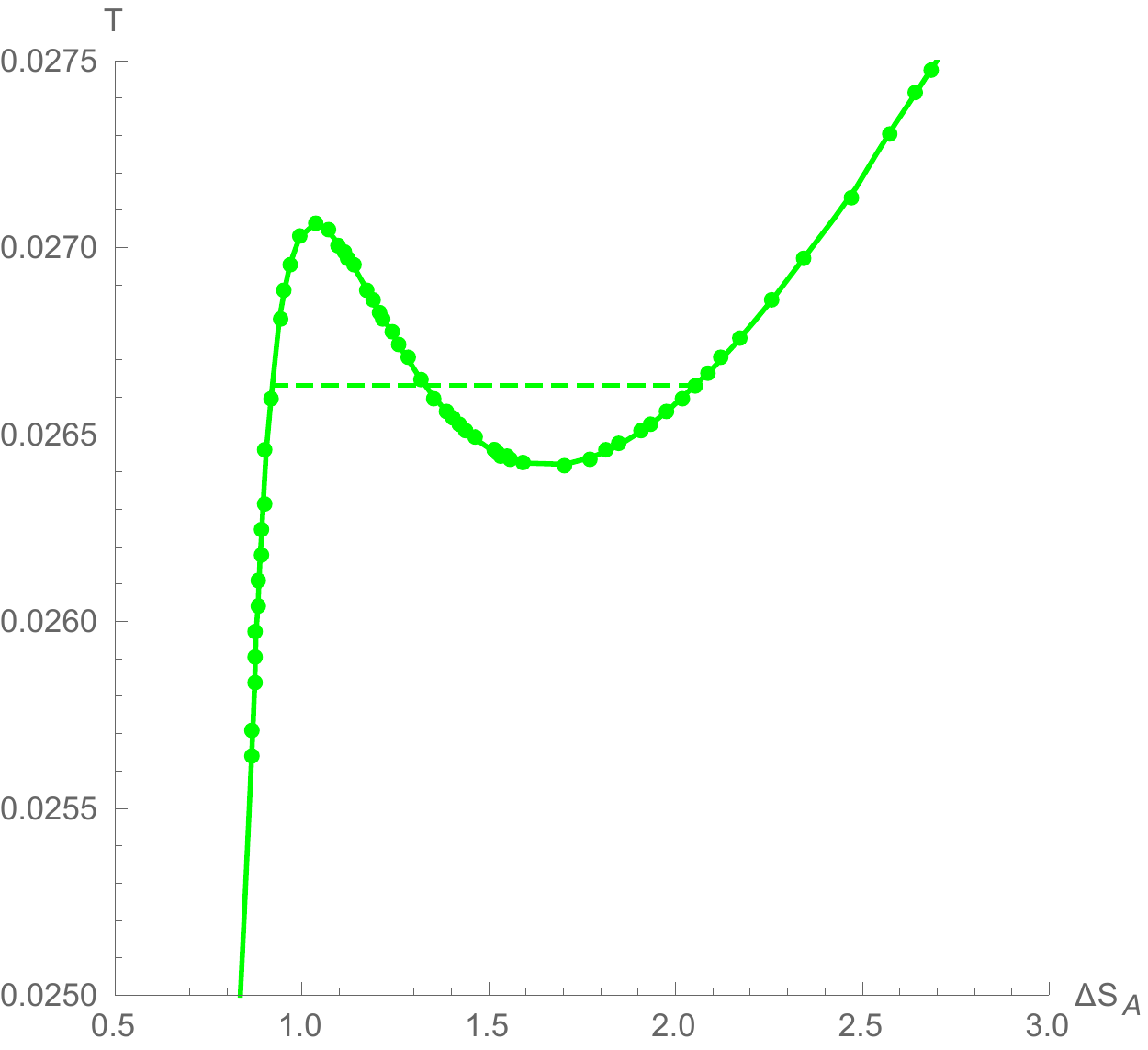}\\
\end{array}
$$
\caption{Plot of isocharges on the $T-\Delta S_{A}$ plane, for $\theta_{0} = 0.1$ (top left), $0.15$ (top right), $0.2$ (bottom left) and $0.4$ (bottom right). For all panels: the values of the charge are $Q=1.5$ (green), $Q=Q_{c}=\frac{10}{6}$ (red) and $Q=1.8$ (orange). The transition isotherm is obtained from the free energy (left panel of Figure \ref{AdSRNPlot}). For the green curves, we also show the data points which were used to create the interpolation.}
\label{AdSRNEEPlot}
\end{figure}

We will take the region $A$ to be a spherical cap on the boundary delimited by $\theta \leq \theta_{0}$. From the remarks in the previous subsection, we will pick four values of $\theta_{0}$ which are smaller than $\frac{\pi}{2}$: $\theta_{0} = 0.1, 0.15, 0.2$ and $0.4$. The minimal surface can be parametrized by the function $r{(\theta)}$, which is independent of the coordinate $\phi$ by rotational symmetry. The area functional is:
\begin{equation}
\mathcal{A} = 2\pi \int_{0}^{\theta_{0}} r\sin{\theta}\sqrt{\frac{(r')^{2}}{f{(r)}}+r^{2}} d\theta\,,
\end{equation}
where $r' \equiv \frac{dr}{d\theta}$. The function $r{(\theta)}$ is the obtained by solving the Euler-Lagrange equation:
\begin{equation}
\frac{\partial L}{\partial r} = \frac{d}{d\theta}\left(\frac{\partial L}{\partial r'}\right)\,,
\end{equation}
with the boundary conditions $r(\theta_{0}) \rightarrow \infty$ and $r'{(0)}=0$ (i.e. the minimal surface is regular at the center $\theta=0$).\\
Also, since EE is UV-divergent, it has to be regularized. We will do it by subtracting the area of the minimal surface in pure AdS whose boundary is also $\theta=\theta_{0}$. In other words, we first integrate the area functional to some cutoff $\theta_{c} \lesssim \theta_{0}$. Then we set $M=Q=0$ to obtain AdS in global coordinates:
\begin{equation}
ds^{2} = -(1+\frac{r^{2}}{L^{2}})dt^{2} + \frac{dr^{2}}{1+\frac{r^{2}}{L^{2}}} + r^{2}d\Omega_{2}^{2}\,.
\end{equation}
The minimal surface which goes to $\theta=\theta_{0}$ on the boundary is given by
\begin{equation}
r_{AdS}{(\theta)} = L\left(\left(\frac{\cos{\theta}}{\cos{\theta_{0}}}\right)^{2}-1\right)^{-1/2}\,.
\end{equation}
We can easily integrate the area functional of this minimal surface up to $\theta_{c}$. Then we subtract this quantity from the black hole one to obtain the renormalized entanglement entropy, which we will denote by $\Delta S_{A}$. If we could compute the area of the minimal surface analytically, then we have to take the limit $\theta_{c} \rightarrow \theta_{0}$ after the subtraction. But since we are doing a purely numerical computation, all we can do is choose $\theta_{c}$ close to $\theta_{0}$.\\
In Figure \ref{AdSRNEEPlot}, we present the plots of the isocharges on the $\Delta S_{A}-T$ plane for the 4 values of $\theta_{0}$. In the case $\theta_{0}=0.1$, we present 3 isocharges corresponding to a charge below the critical one, equal to the critical one, and above the critical one. As can be seen on this plot, the van der Waals phase structure noted in \cite{Johnson:2013dka} is observed. For the 3 other values of $\theta_{0}$, only a subcritical isocharge is presented since this is enough to verify the equal area law. We have also drawn the transition isotherm in dashed green taken from the free energy function.
Finally, we proceed to check whether Maxwell's equal area holds. We will refer to the two areas to be calculated as $A_{1}$ and $A_{2}$:
\begin{equation}
A_{1} = \int_{\Delta S_{A}^{(1)}}^{\Delta S_{A}^{(2)}} T{(\Delta S_{A},Q)} d\Delta S_{A} - T_{*}(\Delta S_{A}^{(2)}-\Delta S_{A}^{(1)})
\end{equation}
\begin{equation}
A_{2} =  T_{*}(\Delta S_{A}^{(3)}-\Delta S_{A}^{(2)}) - \int_{\Delta S_{A}^{(2)}}^{\Delta S_{A}^{(3)}} T{(\Delta S_{A},Q)} d\Delta S_{A}
\end{equation}
where $\Delta S_{A}^{(1)}$, $\Delta S_{A}^{(2)}$ and $\Delta S_{A}^{(3)}$ are the smallest, intermediate, and largest roots of the equation $T_{*} = T{(\Delta S_{A},Q)}$. Maxwell's equal area would amount to the statement:
\begin{equation}
A_{1} = A_{2}
\end{equation}
\begin{table}[t!]
\begin{center}
\begin{tabular}{ |l|l|l|l|l| }
  \hline
  $\theta_{0}$ & $\theta_{c}$ & $A_{1}$ & $A_{2}$ & Relative error  \\ \hline
  0.1 & 0.099 & $1.59197 \times 10^{-6}$ & $1.56443 \times 10^{-6}$ & 1.74 \% \\
  0.15 & 0.149 & $5.40162 \times 10^{-6}$ & $5.30496 \times 10^{-6}$ & 1.8 \% \\
  0.2 & 0.199 & $0.0000127702$ & $0.0000126147$ & 1.22 \% \\
  0.4 & 0.399 & $0.000104962$ & $0.000103373$ & 1.52 \% \\
  \hline
\end{tabular}
\captionof{table}{Comparison of $A_{1}$ and $A_{2}$ for the AdS-RN black hole in 3+1 dimensions.
}
\label{Table1}
\end{center}
\end{table}
We tabulate in table \ref{Table1} the values of $A_{1}$ and $A_{2}$ for each choice of $\theta_{0}$, as well as the choice of the cutoff $\theta_{c}$ for each $\theta_{0}$, and the relative error between $A_{1}$ and $A_{2}$ (taken to be the difference between $A_{1}$ and $A_{2}$ divided by their average). As can be seen from the small relative errors in table \ref{Table1}, it is safe to claim that the equal area law holds for AdS-RN in 4 dimensions.

\section{A counterexample: dyonic AdS-RN}\label{Sec:DyonicAdS4}
In this section, we repeat the procedure above for another charged black hole in AdS: the dyonic solution in 4d, which describes a black hole in AdS carrying both an electric charge and a magnetic charge. This black hole is also a solution to the Einstein-Maxwell action (\ref{EinsteinMaxwell}). The line element together with the gauge field are given by \cite{Caldarelli:2008ze, Lu:2013ura,Dutta:2013dca}:
\begin{equation}
ds^{2} = -f{(r)}dt^{2} + \frac{dr^{2}}{f{(r)}} + r^{2}(d\theta^{2}+\sin^{2}{\theta}d\phi^{2})\,,
\end{equation}
\begin{equation}
A = Q\left(\frac{1}{r_{+}}-\frac{1}{r}\right)dt + P\cos{\theta}d\phi\,,
\end{equation}
with
\begin{equation}
f{(r)} = 1 - \frac{2M}{r} + \frac{Q^{2}+P^{2}}{r^{2}} + \frac{r^{2}}{L^{2}}\,.
\end{equation}
Here $P$ is the magnetic charge.
\subsection{The van der Waals transition of the ``mixed'' ensemble}
\begin{figure}
$$
\begin{array}{cc}
 \includegraphics[width=7.5cm]{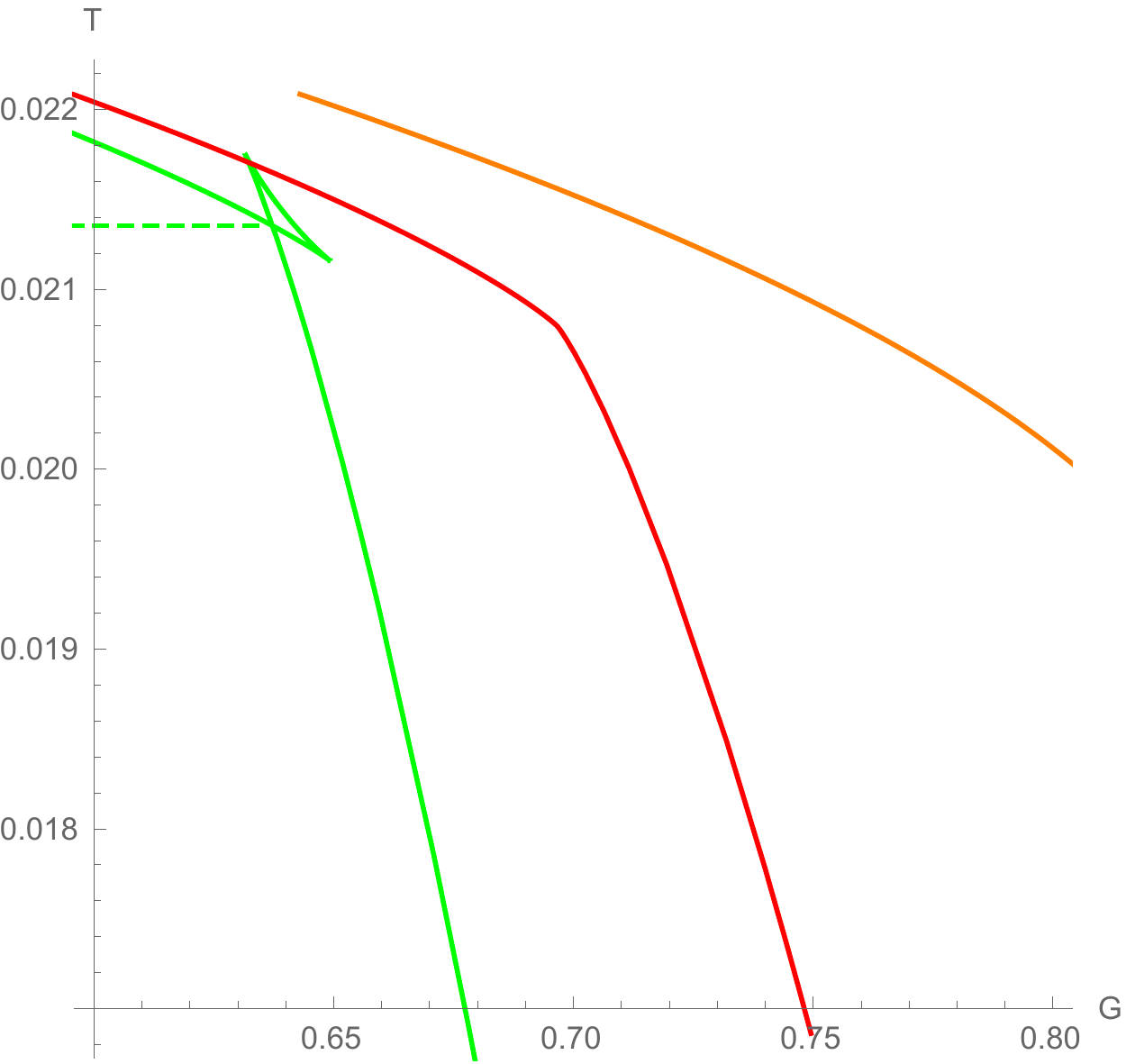} & \includegraphics[width=7.5cm]{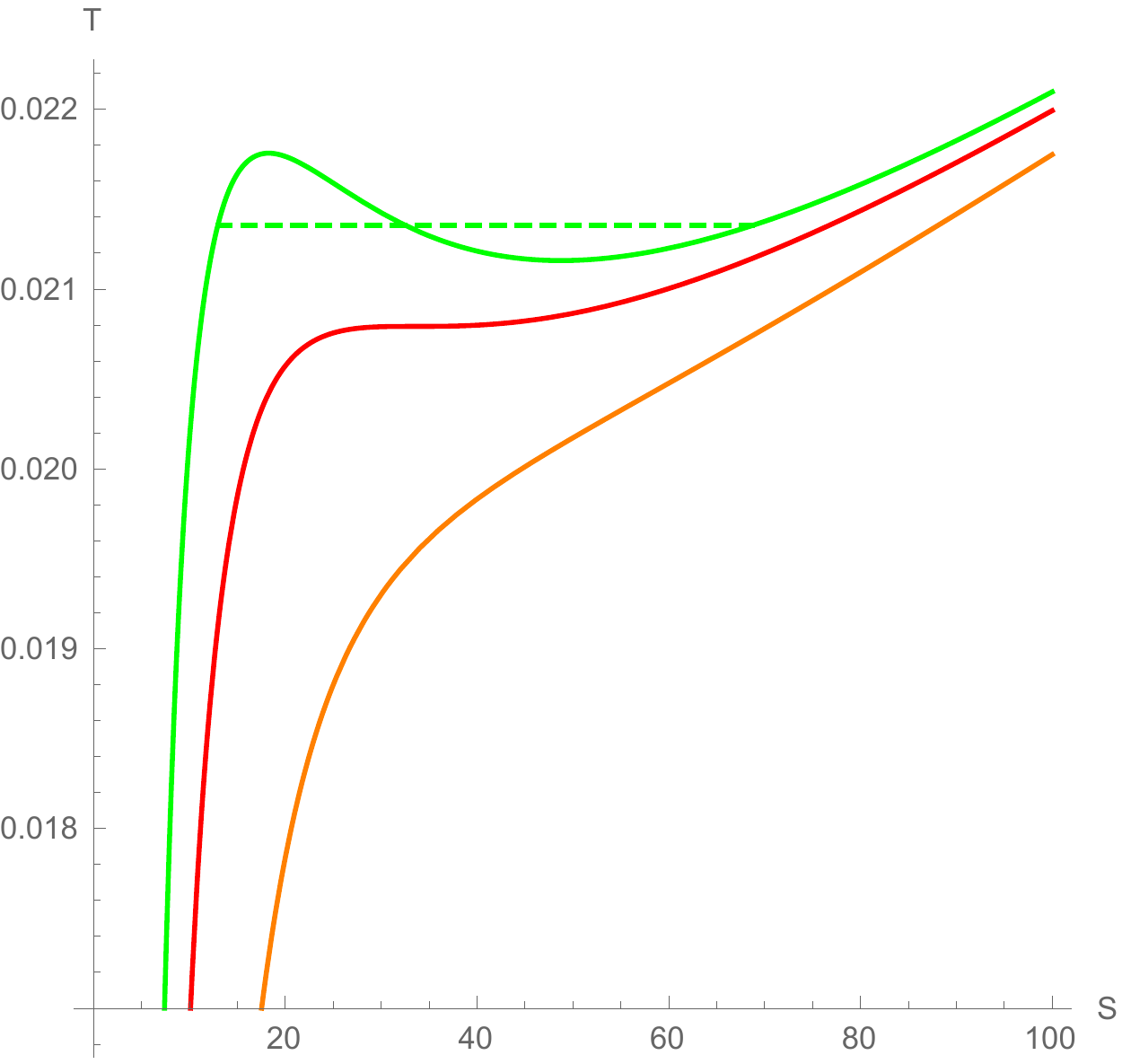}\\
\end{array}
$$
\caption{Plot of the temperature versus the Gibbs potential (left) and the temperature versus the entropy (right) for the dyonic black hole. In both panels, the values chosen for the magnetic charge are $P=0.95$ (green), $P=P_{c}=1.0667$ (red) and $P=1.3$ (orange). We also set $\Phi=0.6$ and $L=10$. The transition temperature for the green curve is $T_{*} = 0.02135$.}
\label{DyonicPlot}
\end{figure}

As for the AdS-RN solution, the phase structure depends on which statistical ensemble one chooses. One can, for example, study the ensemble with both the electric and magnetic charges fixed. This is however not very interesting, since the results in this case can be trivially obtained from the results in section \ref{Sec:AdS4} with the replacement $Q^{2} \rightarrow Q^{2}+P^{2}$.\\
Thus, we will instead work in the fixed $\Phi$, fixed $P$ ensemble. It was observed in \cite{Dutta:2013dca} that the phase structure in this ``mixed'' ensemble exhibits the van der Waals transition. The asymptotic value of the electric potential $\Phi$ is related to the electric charge $Q$ by:
\begin{equation}
\Phi = \frac{Q}{r_{+}}\,.
\end{equation}
The temperature and entropy are given by:
\begin{equation}
T = \frac{1}{4\pi r_{+}}\left(1+\frac{3r_{+}^{2}}{L^{2}}-\Phi^{2}-\frac{P^{2}}{r_{+}^{2}}\right)\,,
\end{equation}
\begin{equation}
S = \pi r_{+}^{2}\,.
\end{equation}
From which we find the function $T(S,\Phi,P)$. Also, since the electric charge is now allowed to vary, we should compute the Gibbs potential:
\begin{equation}
G = M - TS - \Phi Q\,,
\end{equation}
instead of the Helmholtz free energy. Computation gives:
\begin{equation}
G = \frac{3P^{2}}{4r_{+}} + \frac{r_{+}}{4}\left(1-\Phi^{2}-\frac{r_{+}^{2}}{L^{2}}\right)\,.
\end{equation}
If we keep $\Phi$ fixed, and plot the curves of constant $P$ on the $S-T$ plane, we can observe a van der Waals-like transition when $P$ is smaller than a critical value $P_{c}$. The plot of these curves is presented on the right panel of Figure \ref{DyonicPlot}. The critical magnetic charge $P_{c}$, entropy $S_{c}$ and temperature $T_{c}$ can be found to be:
\begin{equation}
P_{c} = \frac{L}{6}(1-\Phi^{2})\,,
\end{equation}
\begin{equation}
S_{c} = \frac{L^{2}}{6}\pi(1-\Phi^{2})\,,
\end{equation}
\begin{equation}
T_{c} = \sqrt{\frac{2}{3}}\frac{1}{L\pi}\sqrt{1-\Phi^{2}}\,.
\end{equation}

We plot the Gibbs potential on the left panel of Figure \ref{DyonicPlot}. From this plot, we obtained the transition temperature $T_{*} = 0.02135$ with $P = 0.95$, $\Phi = 0.6$ and $L=10$. We then checked numerically that $T_{*}$ is indeed the temperature which obeys Maxwell's equal area law: numerical integration of both sides of (\ref{MaxwellLaw}) yields a value of around 0.004658.
\subsection{Maxwell's equal area law for entanglement entropy}
\begin{figure}
$$
\begin{array}{cc}
 \includegraphics[width=8cm]{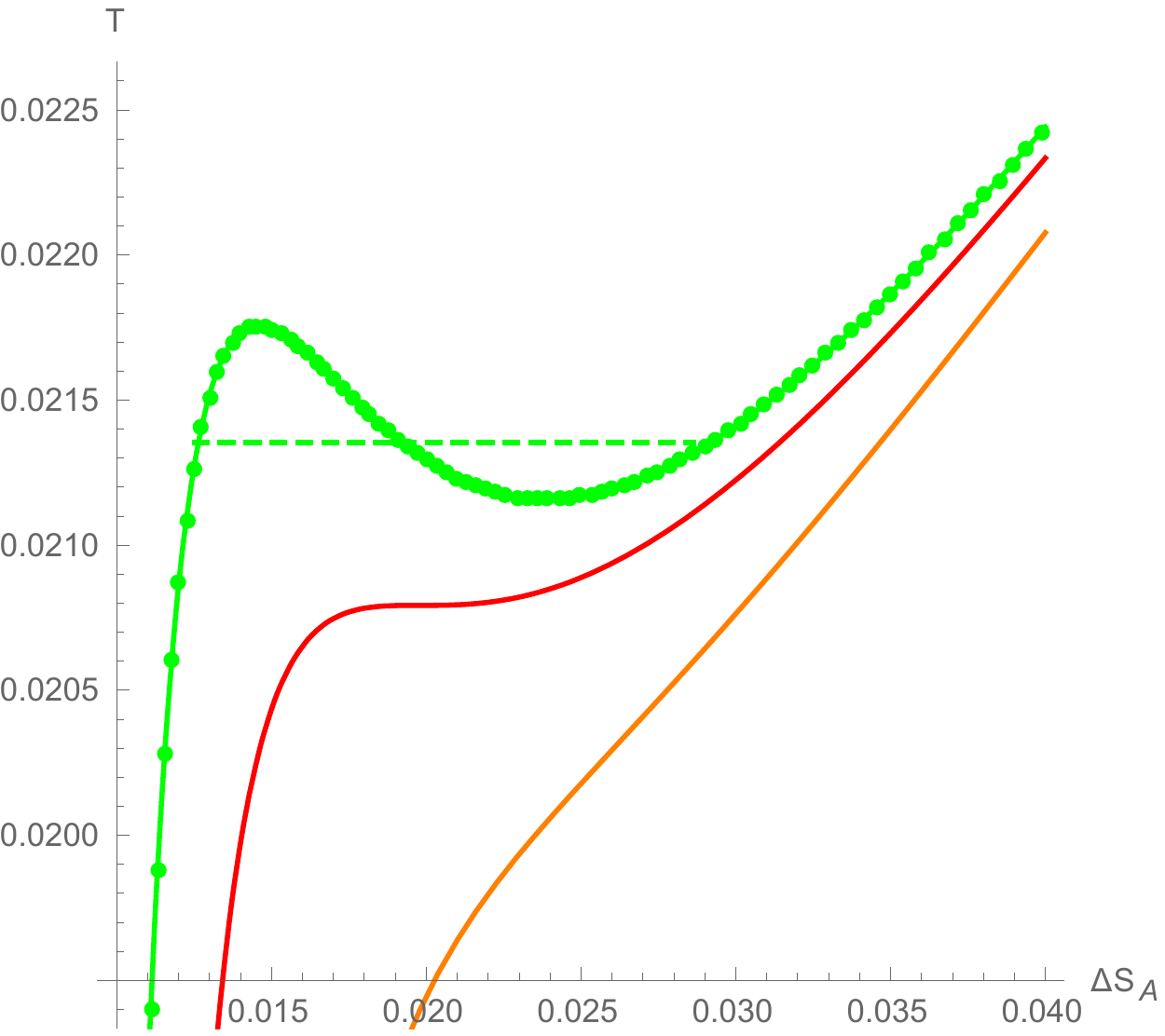} &
 \includegraphics[width=8cm]{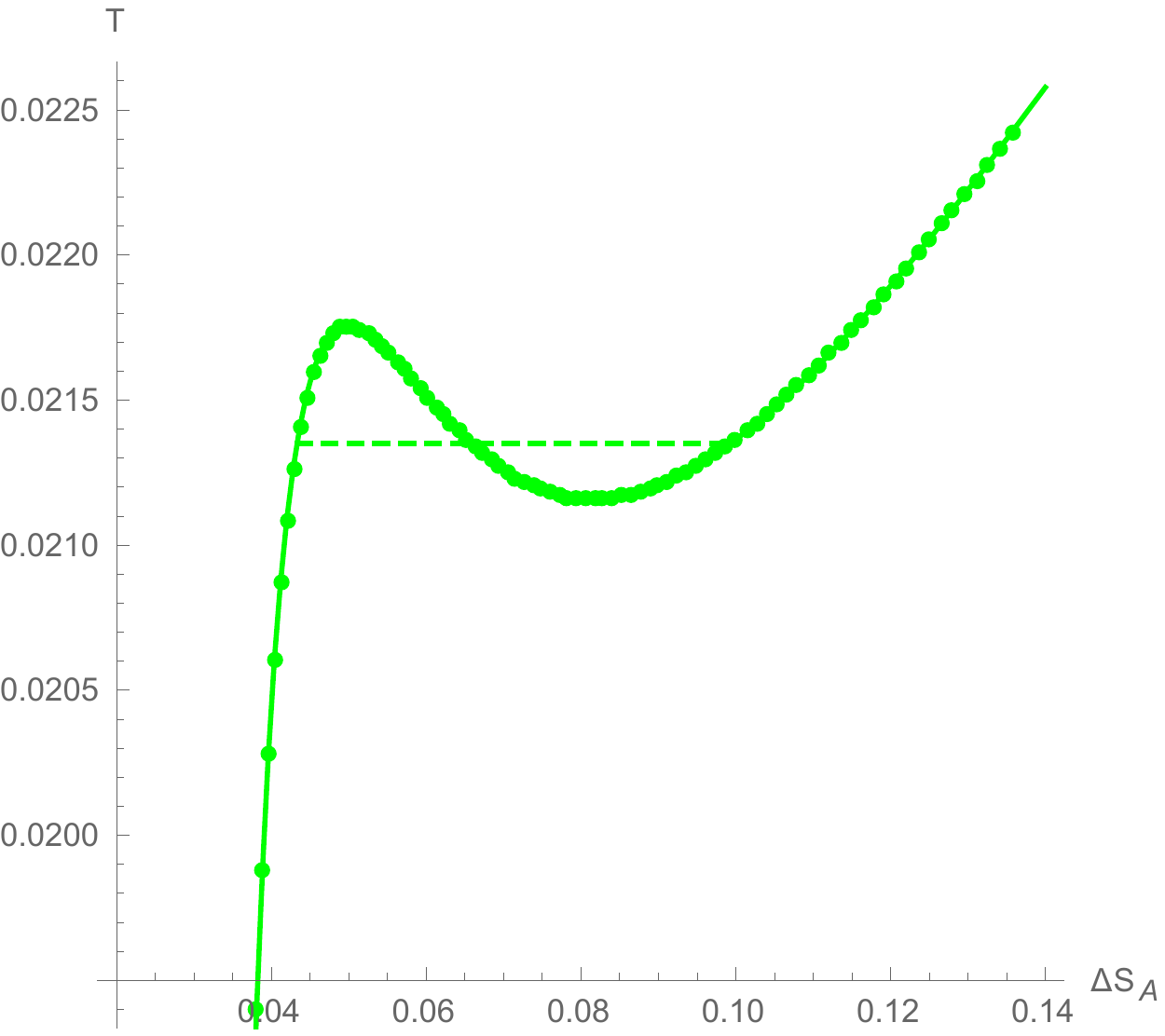} \\
\end{array}
$$
\caption{Plots of curves of constant magnetic charge on the $T-\Delta S_{A}$ plane for the dyonic black hole in the ``mixed'' ensemble, with $\theta_{0}=0.1$ (left) and $0.15$ (right). For both panels, the chosen values of the magnetic charge are $P=0.95$ (green), $P=P_{c}=1.0667$ (red) and $P=1.3$ (orange). We set $\Phi=0.6$ and $L=10$. In both plots, the dashed isotherm is obtained from the Gibbs potential (left panel of Figure \ref{DyonicPlot}). For the green curves, we also show the data points which were used to create the interpolation.}
\label{DyonicEEPlot}
\end{figure}

Next, we turn to the computation of entanglement entropy, and we will show that this time the equal area law is not valid for entanglement entropy, unlike black hole entropy. We will consider two values of $\theta_{0}$: 0.1 and 0.15. The plots of temperature versus entanglement entropy at fixed $P$ are presented in Figure \ref{DyonicEEPlot} (with $\Phi = 0.6$ and $L=10$).
Again, in table \ref{Table2} we tabulate the values of the two areas $A_{1}$ and $A_{2}$ for each choice of $\theta_{0}$, as well as the choice of $\theta_{c}$ and the relative error between $A_{1}$ and $A_{2}$. As can be seen from the large relative errors in table \ref{Table2}, the equal area law does not seem to hold for the dyonic black hole in the mixed ensemble.
\begin{table}[t]
\begin{center}
\begin{tabular}{ |l|l|l|l|l| }
  \hline
  $\theta_{0}$ & $\theta_{c}$ & $A_{1}$ & $A_{2}$ & Relative error \\ \hline
  0.1 & 0.0995 & $1.60959 \times 10^{-6}$ & $1.25275 \times 10^{-6}$ & 24.93 \% \\
  0.15 & 0.1499 & $5.467 \times 10^{-6}$ & $4.2505 \times 10^{-6}$ & 28.62 \% \\
  \hline
\end{tabular}
\captionof{table}{Comparison of $A_{1}$ and $A_{2}$ for the dyonic black hole.
}
\label{Table2}
\end{center}
\end{table}

\section{Another example: 5d AdS-RN}\label{AdS5}
In this section, we will show that the equal area law for entanglement entropy also works for AdS-RN in 4+1 dimensions.

\subsection{The van der Waals transition in the canonical ensemble}
\begin{figure}
$$
\begin{array}{cc}
 \includegraphics[width=7.5cm]{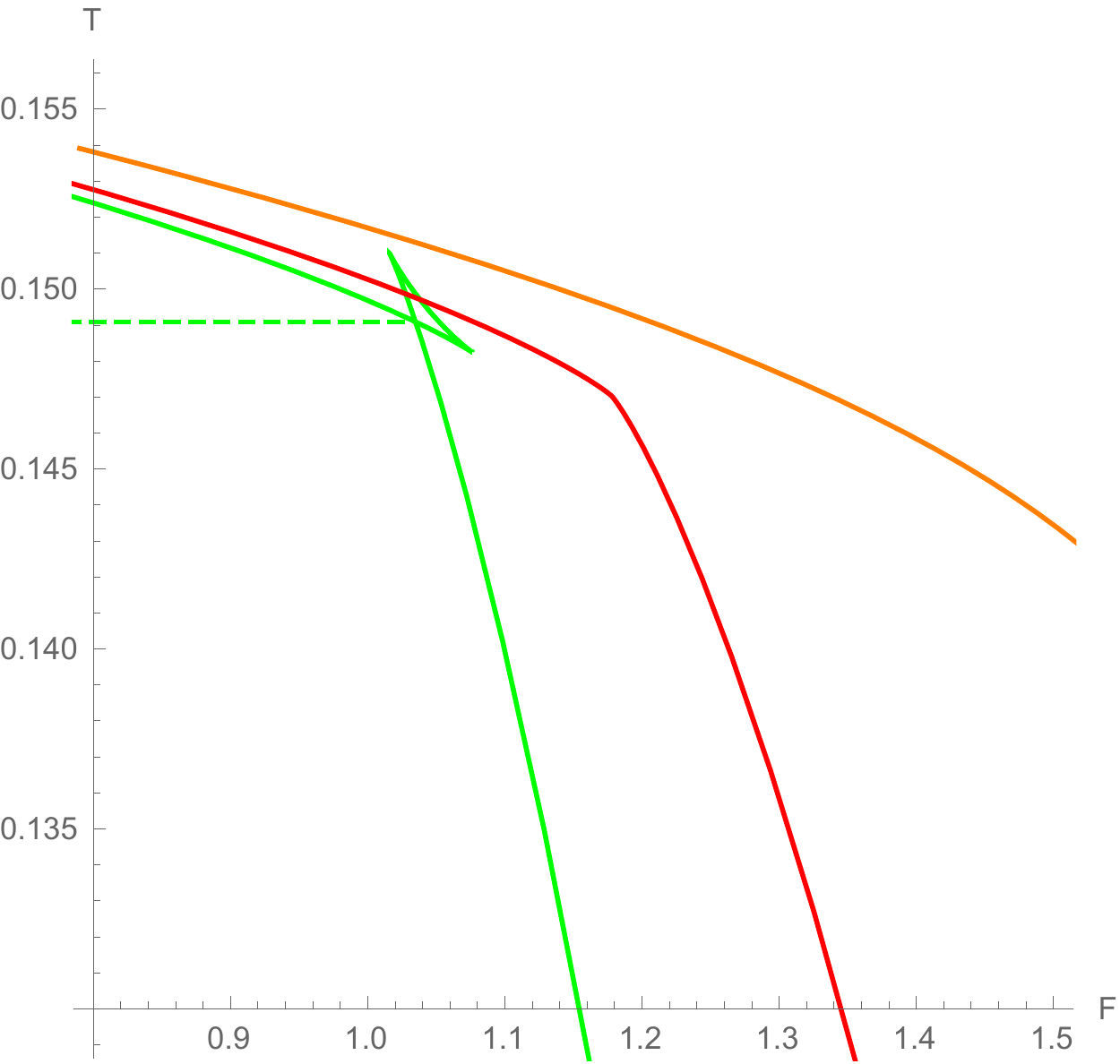} & \includegraphics[width=7.5cm]{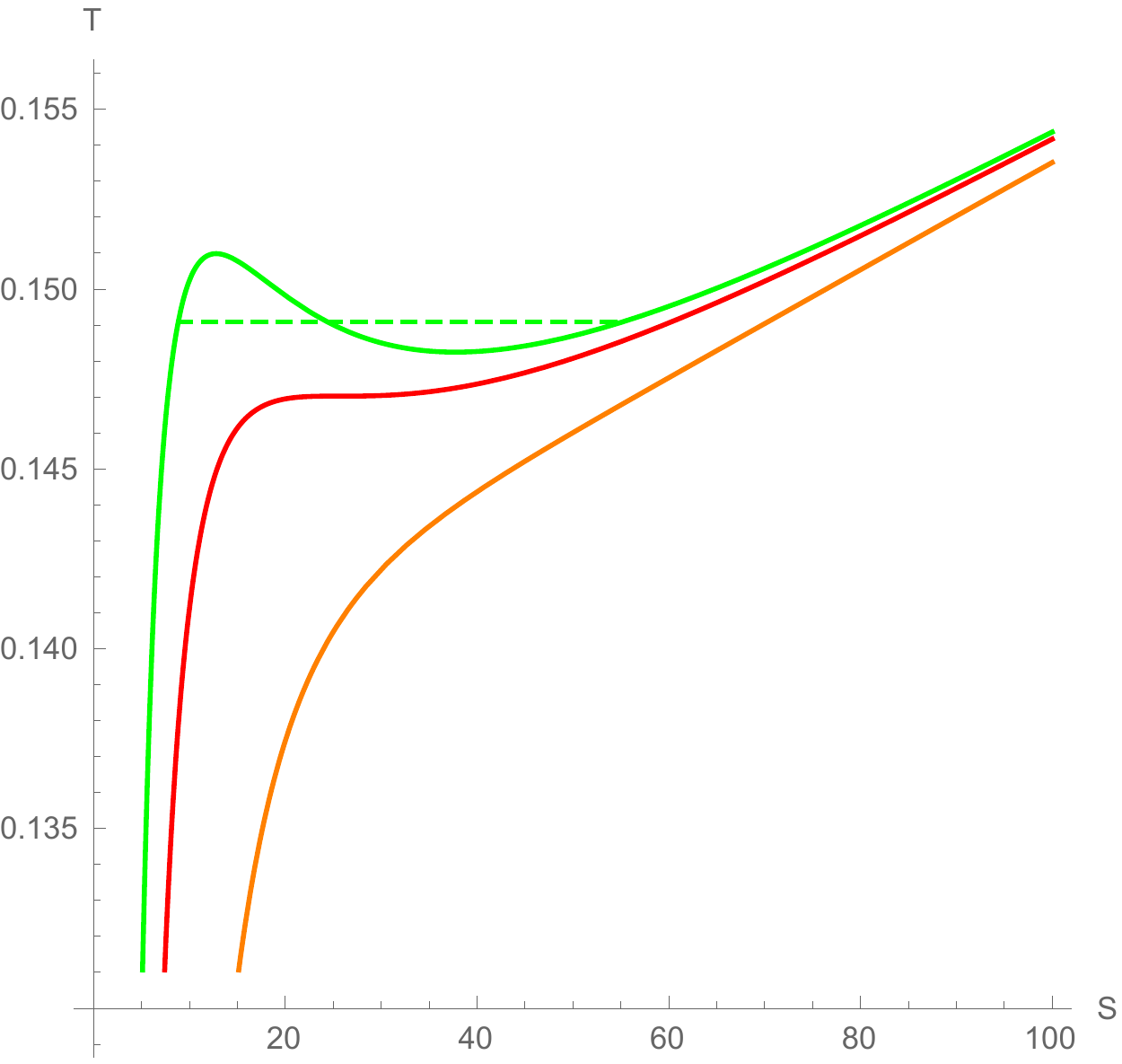}\\
\end{array}
$$
\caption{Left panel: Plot of the free energy versus the temperature for AdS-RN in 4+1 dimensions, with $L=3$ and $Q=1$. Right panel: Plot of isocharges on the $T-S$ plane also with $L=3$ and $Q=1$. The transition temperature for the green curve is $T_{*} = 0.149086$.}
\label{AdSRN5dPlot}
\end{figure}

The AdS-RN solution in arbitrary dimension is given in \cite{Chamblin:1999tk}. For 4+1 dimension, the metric is:
\begin{equation}
ds^{2} = -f{(r)}dt^{2} + \frac{dr^{2}}{f{(r)}} + r^{2}(d\psi^{2} + \sin^{2}{\psi}(d\theta^{2}+\sin^{2}{\theta}d\phi^{2}))
\end{equation}
where $(\psi,\theta,\phi)$ are hyperspherical coordinates on the 3-sphere, with $\psi$ and $\theta$ ranging from 0 to $\pi$, and $\phi$ ranging from $0$ to $2\pi$, and
\begin{equation}
f{(r)} = 1 - \frac{8M}{3\pi r^{2}} + \frac{4Q^{2}}{3\pi^{2}r^{4}} + \frac{r^{2}}{L^{2}}
\end{equation}
The black hole temperature and entropy are given in terms of the horizon $r_{+}$ by:
\begin{equation}
T = \frac{r_{+}}{\pi L^{2}} + \frac{1}{2\pi r_{+}} - \frac{2Q^{2}}{3\pi^{3}r_{+}^{5}}
\end{equation}
\begin{equation}
S = \frac{\pi^{2}}{2}r_{+}^{3}
\end{equation}
From which we obtain the function $T{(S,Q)}$. Following the same steps as the 4-dimensional case, we plot the isocharges on the $T-S$ plane on the right panel of Figure \ref{AdSRN5dPlot}. We find the inflection point to be at:
\begin{equation}
S_{c} = \frac{\pi^{3}}{6\sqrt{3}}L^{3}
\end{equation}
\begin{equation}
Q_{c} = \frac{\pi}{6\sqrt{5}}L^{2}
\end{equation}
\begin{equation}
T_{c} = \frac{4\sqrt{3}}{5\pi L}
\end{equation}
The free energy in the fixed charge ensemble is now:
\begin{equation}
F = \frac{\pi}{8L^{2}}\left(L^{2}r_{+}^{2}-r_{+}^{4}+\frac{20Q^{2}L^{2}}{3\pi^{2}r_{+}^{2}} \right)
\end{equation}
We plot the free energy versus the temperature on the left panel of Figure \ref{AdSRN5dPlot}. The phase structure again presents a van der Waals transition, like in 4d. The transition temperature was found to be $T_{*} = 0.149086$. We can now check numerically the equal area law: evaluating both sides of (\ref{MaxwellLaw}) numerically yields around 0.01711.
\subsection{Maxwell equal area law for entanglement entropy}
Next, we turn to computing the entanglement entropy of a spherical region. The 5d analog of the disk-shaped region $\theta \leq \theta_{0}$ in 4d is the spherical region $\psi \leq \psi_{0}$ for some constant $\psi_{0} \in [0,\pi]$. We will choose 4 different values of $\psi_{0}$: 0.05, 0.1, 0.15, and 0.8. The minimal surface can be parametrized by $r(\psi)$, where this function does not depend on $\theta$ or $\phi$ by rotational symmetry. The area functional is given by:
\begin{equation}
\mathcal{A} = 4\pi \int_{0}^{\psi_{0}} r^{2}\sin^{2}{\psi} \sqrt{r^{2}+\frac{(r')^{2}}{f{(r)}}} d\psi
\end{equation}
Proceeding as for the 4d case, we will solve the equation of motion numerically with the boundary condition that the minimal surface is regular at the center and goes to $\psi_{0}$ at the boundary. To regularize entanglement entropy, we again integrate to some cutoff $\psi_{c}$ and subtract the pure AdS minimal surface which goes to the same $\psi_{0}$ at the boundary. Such a surface is analytically given by:
\begin{equation}
r_{AdS}{(\psi)} = L\left(\left(\frac{\cos{\psi}}{\cos{\psi_{0}}}\right)^{2}-1\right)^{-1/2}
\end{equation}
\begin{figure}
$$
\begin{array}{cc}
 \includegraphics[width=7.5cm]{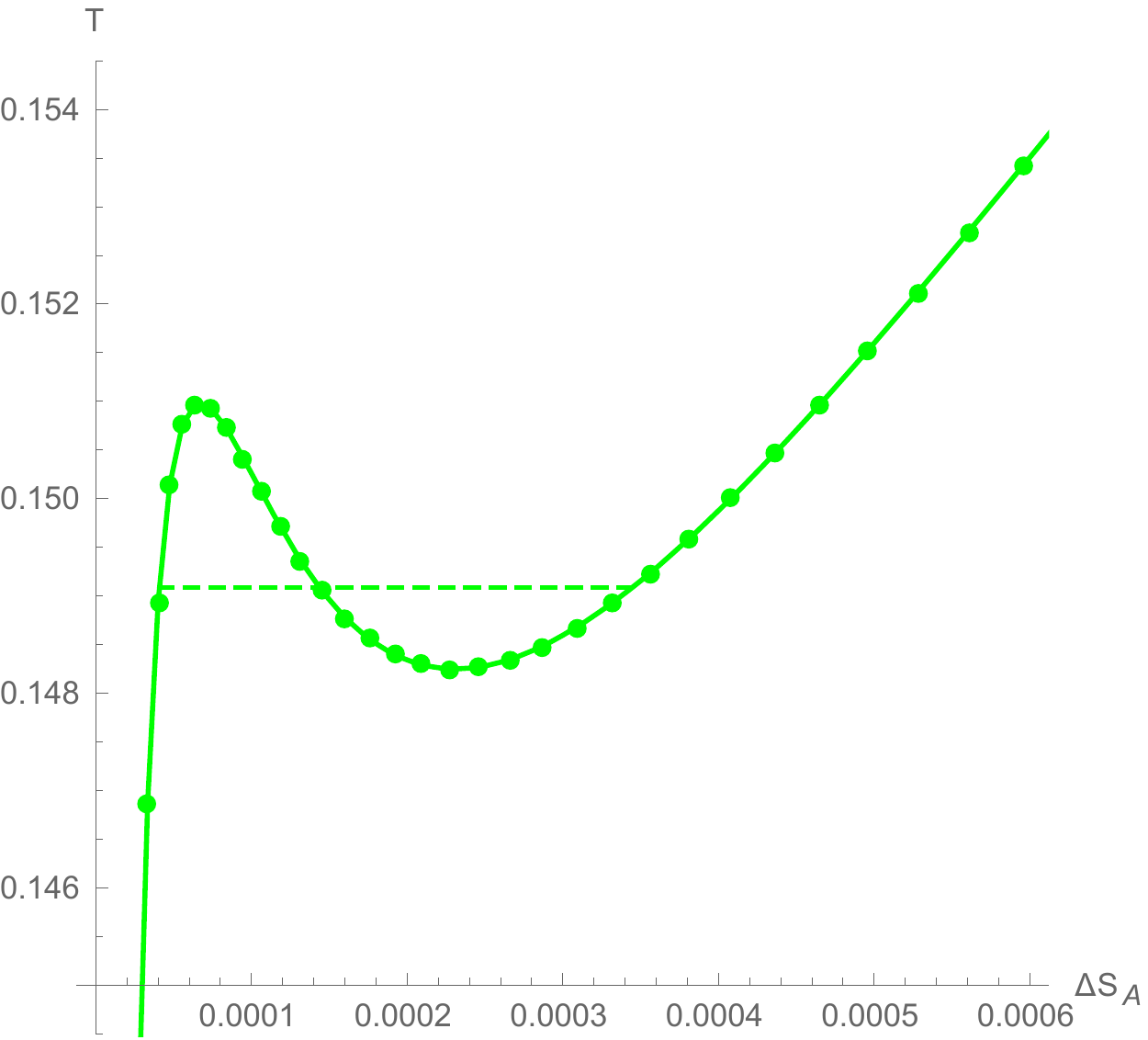} & \includegraphics[width=7.5cm]{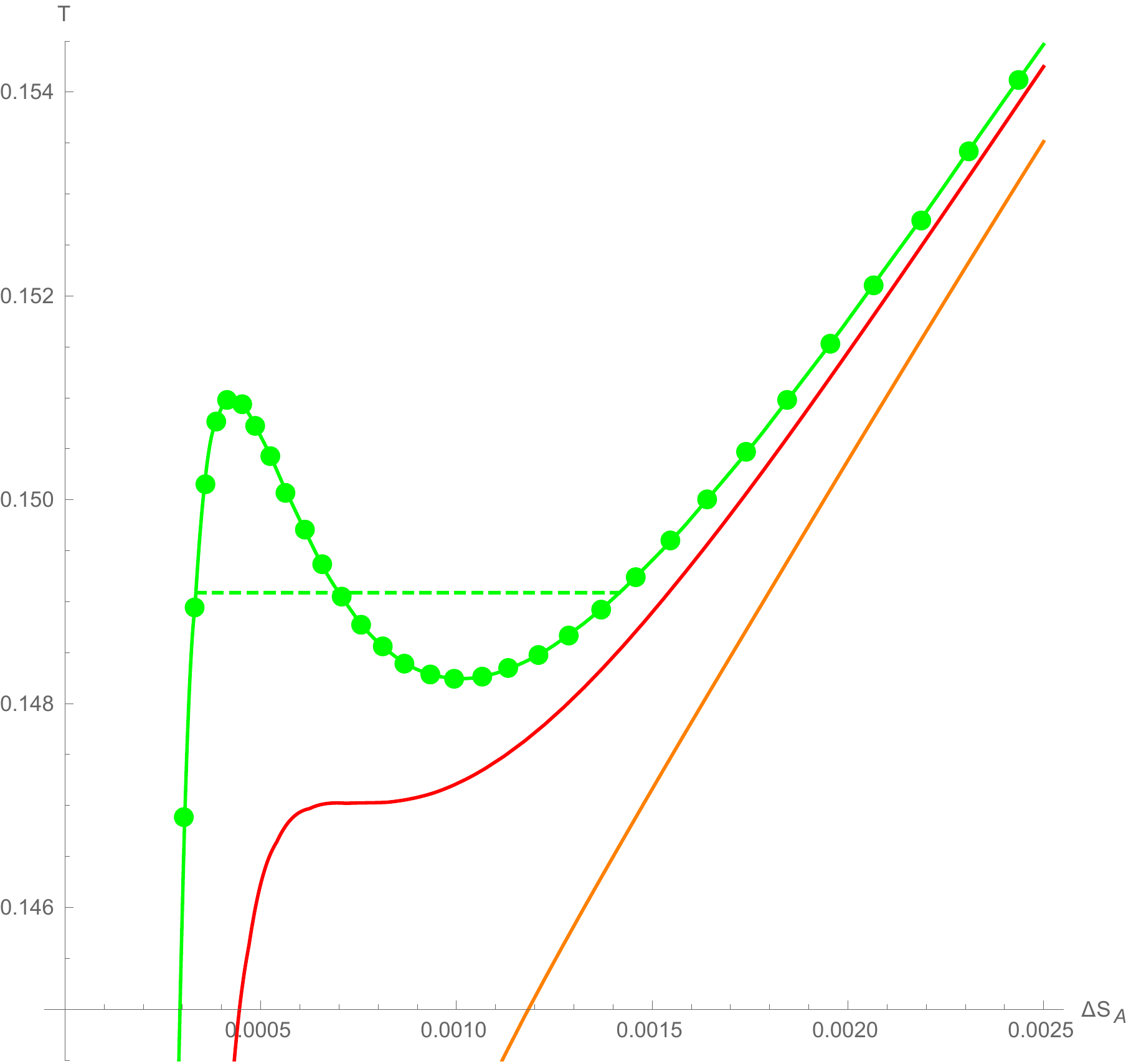} \\
  \includegraphics[width=7.5cm]{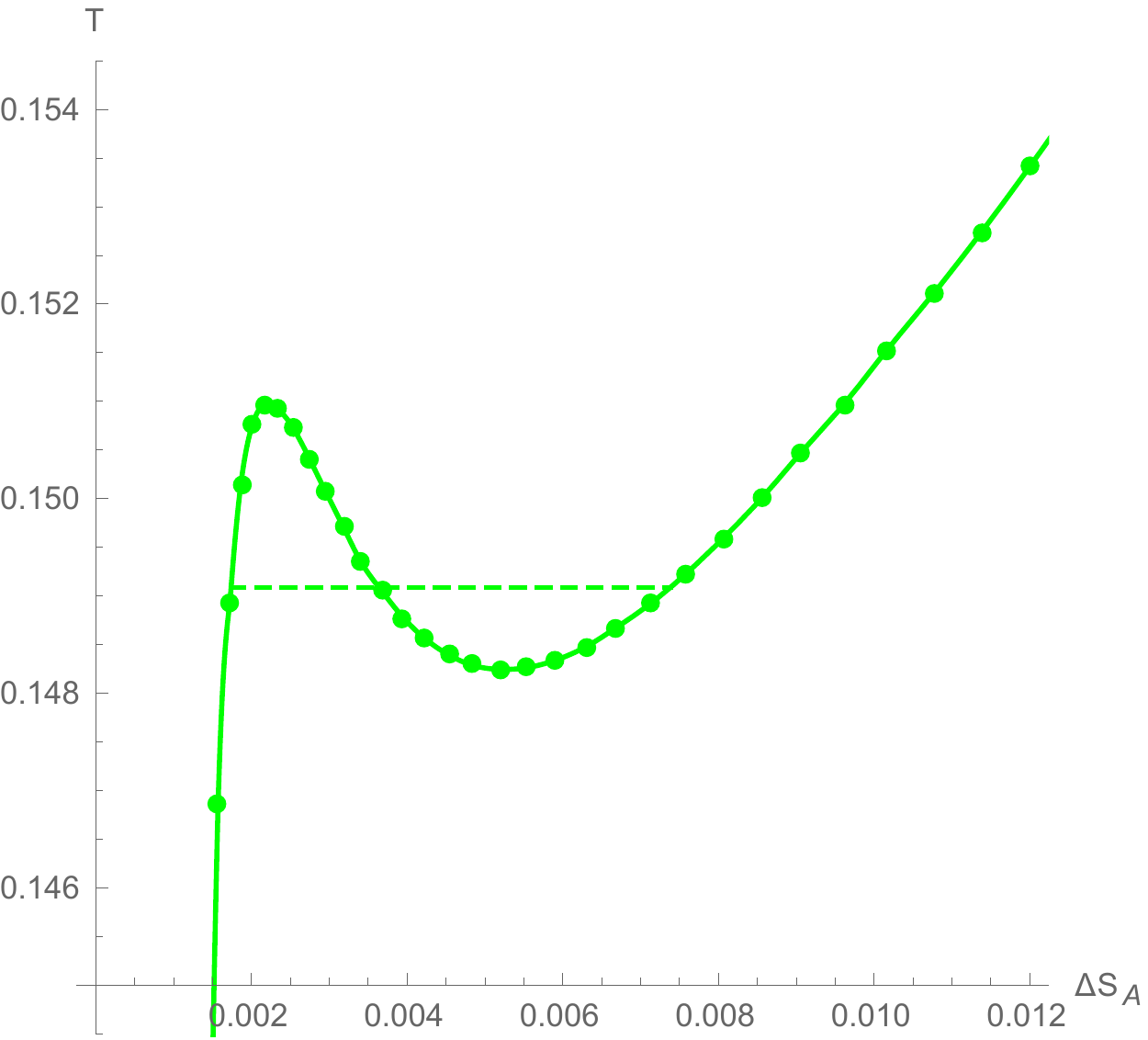} & \includegraphics[width=7.5cm]{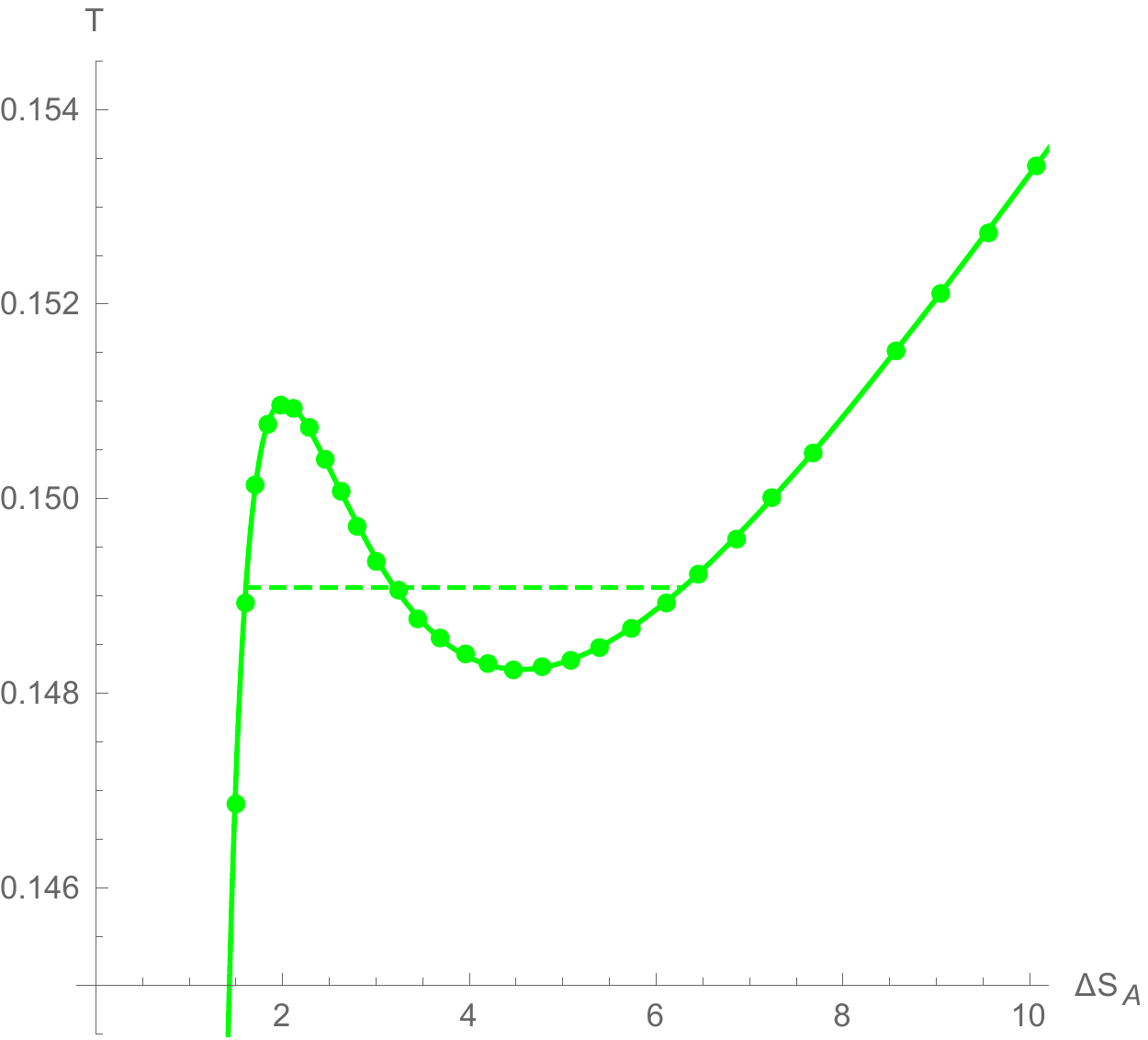} \\
\end{array}
$$
\caption{Plot of isocharges on the $T-\Delta S_{A}$ plane for the 4+1 dimensional AdS-RN solution, and $\psi_{0}=0.05$ (top left), $0.1$ (top right), $0.15$ (bottom left) and $0.8$ (bottom right). The values of the charge chosen are $Q=1.75$ (green), $Q=Q_{c}=\frac{9\pi}{6\sqrt{5}}$ (red) and $Q=3$ (orange). We have set $L=3$. The transition isotherm is obtained from the free energy (left panel of Figure \ref{AdSRN5dPlot}). We also show the data points which were used to create the interpolation.}
\label{AdSRN5dEEPlot}
\end{figure}
We present in Figure \ref{AdSRN5dEEPlot} plots of isocharges for each choice of $\psi_{0}$. As previously, we tabulate in table \ref{Table3} the values of $A_{1}$ and $A_{2}$ for each case, and the relative error between $A_{1}$ and $A_{2}$ and the choice of the cutoff value $\psi_{c}$:
\begin{table}[t]
\begin{center}
\begin{tabular}{ |l|l|l|l|l| }
  \hline
  $\psi_{0}$ & $\psi_{c}$ & $A_{1}$ & $A_{2}$ & Relative error  \\ \hline
  0.05 & 0.0499 & $1.12124 \times 10^{-7}$ & $1.10842 \times 10^{-7}$ & 1.15 \% \\
  0.1 & 0.0999 & $4.05257 \times 10^{-7}$ & $3.98178 \times 10^{-7}$ & 1.76 \% \\
  0.15 & 0.1499 & $2.06799 \times 10^{-6}$ & $2.05255 \times 10^{-6}$ & 0.75 \% \\
  0.8 & 0.799 & $0.00176189$ & $0.00171273$ & 2.87 \% \\
  \hline
\end{tabular}
\captionof{table}{Comparison of $A_{1}$ and $A_{2}$ for the AdS-RN black hole in 4+1 dimensions.
}
\label{Table3}
\end{center}
\end{table}
As in the 4d AdS-RN case, the relative errors are of the order of $1 \%$, and therefore we can safely claim that the equal area law works.

\section{Conclusion}\label{Sec:Conclusion}
The laws of black hole thermodynamics have provided us with a robust understanding of black holes as thermodynamical systems, though the nature of the microscopic constituents of these systems remains by and large mysterious. The gauge-gravity duality seems to suggest, on very general ground, that gravity could emerge from the dynamics of a strongly coupled large-{\it N} quantum field theory. In view of this, it is an important thing to do to see whether observables in such a quantum field theory can mimick the behavior of gravitational systems.

In this paper, we have presented compelling numerical evidence that, while a large class of charged backgrounds in AdS exihibits van der Waals-like phase transitions (in an appropriate statistical ensemble), the transition of the AdS-RN background is special in that its holographic EE obeys the Maxwell's equal area law (for spherical entangling regions). In this respect, we have improved upon previous studies such as \cite{Johnson:2013dka}. Moreover, the equal area law for entanglement entropy seems valid regardless of the size of the entangling region and the dimensionality of the background. That said, we have only considered spherically symmetric entangling regions, and future work to generalize the conclusions in this paper to entangling regions of other shape could yield further interesting insights.

Finally, we note that the van der Waals transition of AdS-RN has been observed in the context of the {\it extended} black hole thermodynamics, where the cosmological constant is identified as the pressure variable and its thermodynamic conjugate as the volume variable \cite{Kubiznak:2012wp}. In this framework, one can analyze the isotherms on the $P-V$ plane, and they turn out to be remarkably similar to the van der Waals gas. One can wonder about the relationship between the van der Waals transition in the $P-V$ plane and the one in the $T-S$ plane. It was pointed in \cite{Spallucci:2013osa} that the two are related to each other by a duality similar to T-duality of string theory. Moreover, it was noted by \cite{Lan:2015bia} that Maxwell's construction in the $P-V$ plane can be surprisingly subtle (the construction does not work if the volume is replaced by the {\it specific volume}, unlike the usual van der Waals gas). We note that the holographic interpretation of the $P-V$ plane is not very well-understood. However, the van der Waals transition on the $P-V$ plane seems to be connected with the renormalization group flow (see \cite{Johnson:2014yja,Caceres:2015vsa}).

Finally, it would be interesting to extend our investigations to other charged black holes in AdS, for example (among others) to more complicated configurations with magnetic field. For instance, there has been lots of recent efforts to construct magnetic stars in AdS (see, among others, \cite{Albash:2012ht, Puletti:2015gwa, Carney:2015dra}) which are the gravitational duals to condensed matter systems. In particular, entanglement entropy has been studied in such a background in \cite{Albash:2015cna}. Even though the equal area law most probably will fail in such backgrounds, the study of other backgrounds may shed light on why it works in the first place for AdS-RN. 

\section{Acknowledgements}
It is a pleasure to thank Elena Caceres, Willy Fischler, Richard Matzner and Juan Pedraza for helpful discussions and comments on the manuscript. We also thank the referee of the previous version of the manuscript for pointing out weak points in the analysis. This research was supported by the National Science Foundation under Grant PHY-1316033.

\end{document}